\documentclass[twocolumn,showpacs,amsmath,amssymb,superscriptaddress,aps,prc]{revtex4-1}
\usepackage[dvips]{graphicx}
\usepackage{epsfig}
\usepackage{amsmath}
\usepackage[dvips]{color}
\newcommand {\beq} {\begin{eqnarray}}
\newcommand {\eeq} {\end{eqnarray}}
\newcommand {\he} {\mbox{$^{6}$He}}
\newcommand {\hebi} {\mbox{$^{6}$He+$^{209}$Bi}}
\newcommand {\heal} {\mbox{$^{6}$He+$^{27}$Al}}
\newcommand {\heni} {\mbox{$^{6}$He+$^{58}$Ni}}
\newcommand {\hesn} {\mbox{$^{6}$He+$^{120}$Sn}}
\newcommand {\hepb} {\mbox{$^{6}$He+$^{208}$Pb}}
\newcommand {\ann} {\mbox{$\alpha +n+n$}}
\newcommand {\emax} {\mbox{$E_{\rm max}$}}
\newcommand {\jmax} {\mbox{$j_{\rm max}$}}
\newcommand {\elab} {\mbox{$E_{\rm lab}$}}
\newcommand {\ecm} {\mbox{$E_{c.m.}$}}
\begin{document}
	\title{Low-energy $\he$ scattering in a microscopic model}
	\author{P. Descouvemont}
	\affiliation{Physique Nucl\'eaire Th\'eorique et Physique Math\'ematique, C.P. 229,
		Universit\'e Libre de Bruxelles (ULB), B 1050 Brussels, Belgium}
\begin{abstract}
A microscopic version of the Continuum Discretized Coupled Channel (CDCC) method is used to investigate $\he$
scattering on $^{27}$Al, $^{58}$Ni, $^{120}$Sn, and $^{208}$Pb at energies around the Coulomb barrier.
The $\he$ nucleus is described by an antisymmetric 6-nucleon wave function, defined in the Resonating Group Method.
The $\he$ continuum is simulated by square-integrable positive-energy states. The model is based only
on well known nucleon-target potentials, and is therefore does not depend on any adjustable
parameter. I show that experimental
elastic cross sections are fairly well reproduced. The calculation suggests that breakup effects increase for
high target masses. For a light system such as $\heal$, breakup effects are small, and a single-channel
approximation provides fair results. This property is explained by a very simple model, based on the
sharp-cut-off approximation for the scattering matrix. I also investigate the $\he$-target optical potentials, which confirm that breakup channels are more and more important when the mass increases. At large distances, polarization effects increase the Coulomb barrier, and provide a long-tail absorption component in the imaginary
part of the nucleus-nucleus interaction.
\end{abstract}
\maketitle

\section{Introduction}

The $\he$ nucleus is the lightest exotic bound system.  Owing to its low separation energy 
($S_{2n}=-0.973$ MeV), $\he$ presents a large radius ($2.33\pm 0.04$ fm) compared to the $\alpha$ 
particle ($1.63\pm 0.03$ fm) \cite{THK92}.  
These observations naturally lead to consider $\he$ as a halo nucleus, where the $\alpha$ core is surrounded by two neutrons \cite{HJJ95}.  This halo structure also exists in other exotic nuclei, such as $^{11}$Li, $^{11}$Be or $^{14}$Be, but the strong binding energy of the $\alpha$ core makes $\he$ a particularly good candidate for precise theoretical models.

Since the 90's, the availability of radioactive beams provided a rich information on the structure and properties of exotic 
nuclei \cite{TSK13,Bo13}. The first experiments essentially focused on reaction cross sections at high energies (see, for example, Ref.\ \cite{THH85}).  
From simple semi-classical models, the radius of the projectile can be deduced from reaction cross sections \cite{SLY03}.  These high-energy experiments have been complemented by breakup cross sections, providing E1 strength distributions.  In most neutron-rich light nuclei, these E1 distributions present a peak at low energies \cite{AN13}.  

More recently, several radioactive-beam experiments were devoted to elastic scattering at low energies, i.e. at energies around the Coulomb barrier.  Typical examples are scattering experiments involving $^{6}$He, $^{8}$Li or $^{11}$Li (see references in Ref.\ \cite{CGD15}).  The main purpose of these data is to provide information on the projectile structure through a reaction model.  In this context, the traditional optical model presents several shortcomings, since the structure of the colliding nuclei is neglected.  Although optical-model calculations may provide some information on the range of the nucleus-nucleus interaction, they are of limited use to derive properties of the projectile.

A significant step forward is provided by the CDCC (Coupled Channel Discretized Continuum) 
method \cite{Ra74b,AIK87}, where the structure of the projectile is taken into account.  The CDCC 
formalism has been initially developed to investigate deuteron-induced reactions \cite{Ra74b}.  
Although the deuteron is not considered as an exotic nucleus, its low binging energy ($B=2.22$ MeV) 
makes breakup channels quite important, even for elastic scattering.  In the CDCC model, the projectile
breakup is simulated by a discrete number of approximate continuum states.  This technique permits a strong 
improvement in the description of deuteron-nucleus cross sections.

In parallel with the development of radioactive beams, the CDCC theory has been abundantly used.  The low binding energy of exotic nuclei makes CDCC an efficient tool, well appropriate to halo nuclei.  Three-body CDCC calculations (i.e.\ where the projectile is defined in terms of two clusters) have been performed on systems involving various projectiles, such as $^{8}$B \cite{LCA09}, $^{11}$Be \cite{DT02} or $^{17}$F \cite{KM12}.  The main steps in CDCC calculations are: (1) the determination of the projectile wave functions, including approximate continuum states, referred to as pseudostates; (2) the calculation of the projectile-target coupling potentials; (3) the resolution of the coupled-channel system; (4) from the scattering matrices and/or from the wave functions, the calculation of the various cross sections (elastic scattering, breakup, fusion, etc).

The application of CDCC to three-body projectiles (i.e., to four-body systems) is, in 
principle, straightforward, as the calculations follow the same procedure as for two-body projectiles.  
In practice, however, CDCC calculations involving three-body projectiles are much more demanding.  
A three-body model for the projectile is obviously more complicated than a two-body model, and the 
level density in the continuum is much higher, leading to coupled-channel systems involving many 
equations (see, for example, Ref.\ \cite{DDC15}).  The first application was performed on the $\hebi$ 
elastic scattering \cite{MHO04}.  Later, other reactions involving $\he$ \cite{MHO04}, $^9$Be \cite{DDC15}
or $^{11}$Li \cite{FCA15} were analyzed with  $\ann$, $\alpha+\alpha+n$ or $^9{\rm Li}+n+n$ descriptions of the projectile.  In most cases, breakup channels play a crucial role to describe elastic scattering.  Even if the ground state wave function of the projectile accurately reproduces the halo structure, single-channel calculations are in general not able to account for the experimental scattering cross sections.

This traditional CDCC approach, where the projectile is described by a two- or by a three-body structure, faces two major problems: (1) for complex projectiles, such as $^{11}$Li, the three-body model is a rather strong
approximation, since it neglects the structure of the core; (2) more important, optical potentials between the target and each constituent of the projectile are often unknown, and crude approximations are sometimes necessary.  These problems have been recently addressed by using a microscopic description of the projectile (MCDCC, see Refs.\ \cite{DH13,DPH15}).  In the MCDCC approach, the projectile wave functions are obtained from a nucleon-nucleon interaction.  To describe the scattering process, only nucleon-target optical potentials are necessary.  These potentials are well known over a broad range of masses and energies.  A first application was performed on the $^{7}$Li system, where it was shown that the MCDCC provides an excellent description of elastic and inelastic scattering, without any adjustable parameter.  In that calculation, the 7-body wave functions of $^{7}$Li are defined in a microscopic $\alpha+t$ cluster approximation, which has been tested on many spectroscopic and scattering properties \cite{Ka86}.

Our aim in the present work is to extend the MCDCC to the $\he$ three-cluster projectile.  In the spirit of Ref.\ \cite{DH13}, I use microscopic $\he$ cluster wave functions, with an exact antisymmetrization between the six nucleons.  The availability of $\he$ microscopic wave functions is recent \cite{KD04,DD09}, and provides an excellent opportunity to improve the theoretical description of $\he$ scattering.
I will consider four systems ($\heal, \heni, \hesn$, and $\hepb$) covering a wide range of target masses, and which have been investigated in various experiments.  The present model offers the possibility 
of a common study with identical conditions of calculations except, of course, in the nucleon-target 
interaction. The $\he$ nucleus is a typical three-body system, and is fairly simple since the core is an
$\alpha$ particle, known to be strongly bound and with a spin $0^+$. General conclusions drawn here
can be, at least partly, extended to other weakly bound three-body systems, such as $^{11}$Li or
$^{14}$Be, which are more difficult to describe in a microscopic approach.

An interesting issue that will be also addressed is the $\he$+target potential.  This has been
discussed in the past \cite{KKR13}, and the conclusions are still controversial \cite{CGD15}.  
It is now accepted that single-channel calculations, using standard $\he$+target potentials, 
are not able to reproduce experimental data on elastic scattering, and that breakup channels 
cannot be neglected.  Deducing equivalent potentials \cite{TNL89}, including breakup effects, 
should bring a valuable information on the nature of the $\he$+target interaction,
and more generally, of the interaction involving exotic nuclei.

In Sec.\ \ref{sec2}, I briefly present the microscopic description of $\he$. Section \ref{sec3} describes
the MCDCC formalism. In Sec.\ \ref{sec4}, I discuss the application to the $\heal$, $\heni$, $\hesn$ and
$\hepb$ elastic scattering. I also try to derive general trends of the $\he$-target interaction, derived from
the MCDCC. In particular, I discuss the role of the halo structure and of breakup in the $\he$ scattering. 
Conclusions and outlook are presented in Sec.\ \ref{sec5}.

\section{Microscopic cluster description of $\he$}
\label{sec2}

The Schr\"odinger equation associated with $\he$ in a partial wave with spin $jm$ and parity $\pi$ reads
\beq
H_0 \, \Psi^{jm\pi}_{(k)}=E^{j\pi}_{(k)} \, \Psi^{jm\pi}_{(k)},
\label{eq1}
\eeq
where label $k$ refers to the excitation level. The 6-body Hamiltonian $H_0$ is given by 
\beq
H_0=\sum_{i=1}^6 T_i +\sum_{i<j=1}^6(V^N_{ij}+V^C_{ij}),
\label{eq2}
\eeq
where $T_i$ is the kinetic energy of nucleon $i$, and $V^N_{ij}$ and $V^C_{ij}$ are the nuclear and Coulomb interactions between nucleons $i$ and $j$.  The nuclear term is taken as the Minnesota potential \cite{TLT77}, involving the exchange parameter $u$, and complemented by a zero-range spin-orbit force \cite{DD12}.

Equation (\ref{eq1}) is solved by using the cluster approximation. In other words, the $\he$ nucleus is represented by
a six-body wave function, but approximated by an $\alpha$ core and two neutrons.  This leads to the Resonating 
Group Method (RGM, see Refs.\ \cite{Ho77,DD12}) wave function
\begin{align}
\label{eq3}
\Psi^{jm\pi}_{(k)}=& {\cal A} \,
\sum_{\gamma}\sum_{K=0}^{\infty} \phi_{\alpha}
\biggl[ \bigl[\phi_n \otimes \phi_n \bigr]^S 
\otimes Y^{\ell}_{\ell_x \ell_y K}(\Omega_{\rho})\biggr]^{jm}\nonumber \\
& \times \chi^{j\pi}_{(k)\gamma K}(\rho),
\end{align}
where I use the hyperspherical formalism with $\rho$ as hyperradius \cite{ZDF93,KD04}.  
In Eq.\ (\ref{eq3}), ${\cal A}$ is the six-nucleon antisymmetrizor, 
$\phi_{\alpha}$ is a $(0s)^4$ shell-model wave function of the $\alpha$ particle, and $\phi_n$ is a spinor associated with the neutrons. 
The total spin 
$S=0,1$ results from the coupling of the neutron spins, and the total angular momentum $\ell$ from the coupling of the angular momenta
$\ell_x$ and $\ell_y$, associated with the Jacobi coordinates $\pmb{x}$ and $\pmb{y}$. 
Index $\gamma$ stands for $\gamma=(\ell_x,\ell_y,\ell,S)$, and the hypermoment $K$ runs from zero to infinity.  In practice a truncation value $K_{\rm max}$ is adopted.  The hyperspherical functions $Y^{\ell}_{\ell_x \ell_y K}(\Omega_{\rho})$ are well known (see, for example, 
Ref.\ \cite{ZDF93}), and depend on five angles $\Omega_{\rho}=(\Omega_{x},\Omega_{y},\alpha)$, where
$\alpha$ is the hyperangle.
The hyperradial functions $\chi^{j\pi}_{(k)\gamma K}(\rho)$ are to be determined from the Schr\"odinger equation (\ref{eq1}).

As for two-cluster systems, the RGM definition clearly displays the physical interpretation of the cluster approximation.  In practice, however, using the Generator Coordinate Method (GCM) wave functions is equivalent, and is more appropriate to systematic numerical calculations \cite{DD12}.  In the GCM, the wave function (\ref{eq3}) is equivalently written as  
\beq
\Psi^{jm\pi}_{(k)}=\sum_{\gamma,K} \int dR \, f^{j\pi}_{(k)\gamma K}(R)
\, \Phi^{jm\pi}_{\gamma K}(R),
\label{eq4}
\eeq
where $R$ is the generator coordinate, $\Phi^{jm\pi}_{\gamma K}(R)$ are projected Slater determinants, 
and $f^{j\pi}_{(k) \gamma K}(R)$ are the generator functions (see Ref.\ \cite{DD12} for more detail).  
In practice, the integral is replaced by a sum over a finite set of $R$ values (typically 10 values are chosen).  

After discretization of (\ref{eq4}), the generator functions are obtained from the eigenvalue problem,
known as the Hill-Wheeler equation,
\begin{align}
\sum_{\gamma K n} & \biggl[
H^{j\pi}_{\gamma K,\gamma' K'}(R_n,R_{n'})
- E^{j\pi}_{(k)} N^{j\pi}_{\gamma K,\gamma' K'}(R_n,R_{n'})\biggr] \nonumber \\
& \times
 f^{j\pi}_{(k) \gamma K}(R_n)=0,
\label{eq5}
\end{align}
where the Hamiltonian and overlap kernels are obtained from 7-dimension integrals involving matrix elements between Slater determinants.  
These matrix elements are computed with Brink's formula \cite{Br66}, and the main part of the numerical calculations is devoted to the multidimension integrals (see Refs.\ \cite{KD04,DD12} for detail).

In addition to the overlap and Hamiltonian kernels
\beq
N^{j\pi}_{\gamma K,\gamma' K'}(R_n,R_{n'})&=& 
\langle \Phi^{jm\pi}_{\gamma K}(R_n) \vert \Phi^{jm\pi}_{\gamma' K'}(R_{n'}) \rangle \nonumber \\
H^{j\pi}_{\gamma K,\gamma' K'}(R_n,R_{n'})&=& 
\langle \Phi^{jm\pi}_{\gamma K}(R_n) \vert H \vert \Phi^{jm\pi}_{\gamma' K'}(R_{n'}) \rangle,
\label{eq6}
\eeq
I also need matrix elements of the densities
\beq
 & &\rho^{jm\pi,j'm'\pi'}_{\gamma K,\gamma' K'}(\pmb{r},R_n,R_{n'})= \nonumber \\
 & &\langle \Phi^{jm\pi}_{\gamma,K}(R_n) \vert 
\sum_i \bigl(\frac{1}{2} \pm t_{iz}\bigr)\delta(\pmb{r}-\pmb{r}_i)
\vert \Phi^{jm\pi}_{\gamma' K'}(R_{n'}) \rangle,
\label{eq7}
\eeq
where $\pmb{t}_i$ is the isospin of nucleon $i$, and where the $+$ and $-$ signs correspond to the neutron and proton densities, respectively.  These matrix elements are computed as explained 
in Ref.\ \cite{BDT94}.  The proton and neutron densities are defined as
\beq
 & &\rho^{jm\pi,j'm'\pi'}_{k,k'}(\pmb{r}) \nonumber \\
&  &=\langle \Psi^{jm\pi}_{(k)} \vert 
\sum_i \bigl(\frac{1}{2} \pm t_{iz}\bigr)\delta(\pmb{r}-\pmb{r}_i)
\vert \Psi^{j'm'\pi'}_{(k')} \rangle ,
\label{eq8}
\eeq
and are determined from the generator functions and from the matrix elements (\ref{eq7}). The sign "$-$"
correspond to proton ($p$) density, and the sign "$+$" to the neutron ($n$) density. 
The densities are necessary to compute the $\he$-target coupling potentials (see Sec. \ref{sec3}). 
They are expanded in multipoles \cite{BDT94,Ka81} as
\beq
& &\rho^{jm\pi,j'm'\pi'}_{k,k'}(\pmb{r}) =\sum_{\lambda}
\langle j'\, m'\, \lambda \, m-m' \vert j \, m \rangle \nonumber \\
&& \times Y_{\lambda}^{m-m' \star}(\Omega_r) \, \rho^{j\pi,j'\pi'}_{\lambda (k,k')}(r),
\label{eq8a}
\eeq
and the normalization is such that
\beq
\int \rho^{jm\pi,jm\pi}_{k,k}(\pmb{r})\, d\pmb{r}=Z {\rm \ or\ }N. 
\label{eq8b}
\eeq
For the sake of clarity in the notations, I do not explicitly write indices $p$ and $n$ for the proton and
neutron densities, respectively.

As usual in CDCC calculations, the continuum of the projectile is simulated by positive-energy wave functions, referred to as pseudostates.  In other words, $k$ values corresponding to $E^{j\pi}_{(k)}<0$ are physical states (for $\he$ only the ground state is bound), and $k$ values corresponding to $E^{j\pi}_{(k)}>0$ are associated with pseudostates (or with narrow resonances).

\section{Outline of the MCDCC}
\label{sec3}

The MCDCC has been presented in Refs.\ \cite{DH13,DPH15} for two-cluster projectiles.  I give here a brief outline, by emphasizing specificities of three-cluster projectiles.
For a system involving a projectile associated with $H_0$ [see Eq.\ (\ref{eq1})], the total Hamiltonian is defined by
\beq
H=H_0(\pmb{r}_i)+T_R+\sum_i v_{iT}(\pmb{r}_i- \pmb{R}),
\label{eq9}
\eeq
where $\pmb{r}_i$ are the internal coordinates of the projectile, and $ \pmb{R}$ is the projectile-target relative coordinate.  
In the isospin formalism, the interaction between nucleon $i$ and the target $T$ reads
\beq
v_{iT}(\pmb{s})&=&\bigl(\frac{1}{2} - t_{iz}\bigr) \biggl[v_{pT}(\pmb{s})+\frac{Z_Te^2}{s}\biggr] \nonumber \\
&&+ \bigl(\frac{1}{2} + t_{iz}\bigr) v_{nT}(\pmb{s}),
\label{eq10}
\eeq
where $Z_Te$ is the charge of the target,  and $v_{pT}(\pmb{s})$ 
and $v_{nT}(\pmb{s})$ are proton and neutron optical potentials, respectively.  Their imaginary parts simulate the excitation of the target.  

In the CDCC formalism, the total wave function, associated with Hamiltonian (\ref{eq9}) is expanded as 
\beq
\Psi^{JM\pi}(R)=\sum_{j kL}\,
\varphi^{JM\pi}_{j kL}(\Omega_R,\pmb{r}_i)\,
g^{J\pi}_{j kL}(R),
\label{eq11}
\eeq
where $J$ and $\pi$ are the total angular momentum and parity, respectively. The channel functions are defined by
\beq
\varphi^{JM\pi}_{j kL}(\Omega_R,\pmb{r}_i)=i^L
\biggl[\Psi^{j}_{(k)}(\pmb{r}_i)\otimes
Y_L(\Omega_R)
\biggr]^{JM}.
\label{eq12}
\eeq

In Eq.\ (\ref{eq11}), the sums over $k$ and $j$ are truncated at a maximum energy $\emax$, and at a maximum angular momentum $\jmax$, respectively (notice that I assume that the parity of the projectile is implied in $j$). 
The radial functions $g^{J\pi}_{c}(R)$ (I use $c=(j, k,L)$) are given by the coupled-channel system
\beq
\bigl( T_L +E_{c}-E \bigr) 
g^{J\pi}_c(R) +\sum_{c'}V^{J\pi}_{c,c'}(R) g^{J\pi}_{c'}(R)=0,
\label{eq13}
\eeq
where $E_c$ are the threshold energies, and where the
kinetic operator is defined by
\beq
T_L=-\frac{\hbar^2}{2\mu_{PT}}\left( \frac{d^2}{dR^2}-\frac{L(L+1)}{R^2}\right),
\label{eq13b}
\eeq
$\mu_{PT}$ being the reduced mass of the system.
 
The coupling potentials are given by the matrix elements
\beq
V^{J\pi}_{c,c'}(R)=\langle \varphi^{JM\pi}_{c} \vert \sum_i v_{iT}(\pmb{r}_i- \pmb{R}) \vert 
\varphi^{JM\pi}_{c'} \rangle.
\label{eq14}
\eeq
The calculation is performed by Fourier transforms of the nucleon-target interaction, and of the GCM densities \cite{SL79}.  Technical details are provided in Ref.\ \cite{GD14}.

At large distances, i.e.\ when only the monopole Coulomb interaction remains in (\ref{eq14}), 
the solutions 
of the coupled-channel system (\ref{eq13}) are given by
\begin{eqnarray}
g^{J\pi}_{c}(R)\rightarrow
\left\{\begin{array}{ll}
v_{c}^{-1/2} \Bigl( I_{L} (k_{c} r)\delta_{c \omega} -  O_{L} (k_{c} r)U^{J\pi}_{c \omega} \Bigr),
&E> E_c, \\
A_{c}^{\omega} W_{-\eta_{c},L+1/2}(2k_{c}r), & E<E_c .
\end{array} \right .
\label{eq14b}
\end{eqnarray}
In these definitions, $v_c$ and $k_c$ are the velocity and wave number in channel 
$c$, and $\omega$ is the entrance channel.  
Functions $I_{L}(x)$ and $O_{L}(x)$ are the incoming and outgoing Coulomb functions \cite{Th10}, 
and $W_{a,b}(x)$ is the Whittaker function \cite{Ol10}.  Equations (\ref{eq14b}) define the scattering 
matrices $\pmb{U}^{J\pi}$, which are used to compute cross sections \cite{Th88,CH13}.  System (\ref{eq13}) is 
solved by the $R$-matrix technique, using a Lagrange basis \cite{DB10,De15}.  This method 
represents a useful tool to determine the scattering matrices, even for many-channel calculations.  
The elastic cross sections are then determined from standard formula \cite{CH13}.

\section{Application to $\he$ scattering}
\label{sec4}

\subsection{Conditions of the calculation}
In the CDCC method, the first step before the cross section calculations is the determination of $\he$ wave functions (\ref{eq3}, \ref{eq4}).  I take $N=8$ values for the generator coordinate associated with the hyperradius $R$ ($R=1.5$ fm to 12 fm by step of 1.5 fm).  The parameters of the Minnesota interaction are $u=1.0045$ and $S_0=37$ MeV.fm$^5$ which reproduce the $\he$ binding energy ($-0.973$ MeV) and 
the $\alpha+n$ phase shifts \cite{TMO07}.  The oscillator parameter is chosen as $b=1.36$ fm, a standard value for the $\alpha$ particle.

With these conditions, the matter and charge radii are computed as $\sqrt{<r^2>_m}=2.35$ fm and $\sqrt{<r^2>_p}=1.80$ fm. The matter radius is in excellent agreement with experiment ($2.33\pm 0.04$ fm \cite{THK92}). The charge radius
has been measured with a high accuracy ($2.054 \pm 0.014$ fm), by using laser spectroscopy \cite{WMB04}. The RGM, as most cluster theories (see the discussion in Ref.\ \cite{WMB04}) slightly underestimates this value. Most likely the $t+t$ configuration might play a role to explain
the experimental charge radius.

\begin{figure}[htb]
	\begin{center}
		\epsfig{file=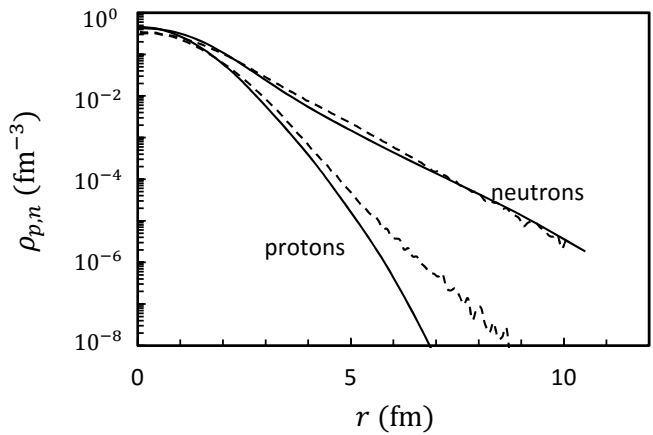,width=8.6cm}
		\caption{Proton ($\rho_p$) and neutron ($\rho_n$) monopole densities of the $\he$ ground state ($j\pi=0^+,k=1$ in 
			Eq.\ (\ref{eq8a})). The dashed lines represent densities
			obtained with the {\sl ab initio} model of Ref.\ \cite{PPC97}.}
		\label{fig_dens}
	\end{center}
\end{figure}

The proton and neutron monopole densities of the ground state 
are shown in Fig.\ \ref{fig_dens} (for a spin $j=0$, only the monopole term $\lambda=0$ contributes to the expansion (\ref{eq8b})).  
As expected, the neutron density presents a slow decrease at large distance, in agreement with the picture of a "neutron halo".  For comparison I also present in Fig.\ \ref{fig_dens} 
the densities obtained in the Green's-function Monte Carlo method with the Argonne $v_{18}$ interaction \cite{PPC97}.
The goal of the present work is not to focus on an optimal description of $\he$. However, I have here
a good opportunity to compare the densities of the ground state with those of an {\sl ab initio} model.
At large distances, the proton densities are slightly lower in the present model, as expected from the cluster 
approximation. However the neutron densities, accurately described by a cluster model, are very close to each other.

\begin{figure}[htb]
	\begin{center}
		\epsfig{file=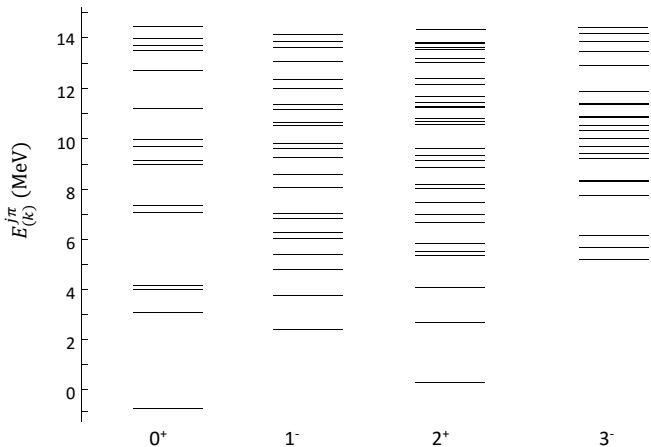,width=8.6cm}
		\caption{$\he$ pseudostates for $j=0^+ - 3^-$. Energies are defined from the $\alpha + n +n$ threshold.}
		\label{fig_spec}
	\end{center}
\end{figure}

The $\he$ spectrum for the $j=0^+ - 3^-$ partial waves is shown in Fig.\ \ref{fig_spec} up to 15 MeV.  The only bound state is the $0^+$ ground state.  
The $2^+$ narrow resonance is predicted at an energy lower than experimentally, as already observed in previous calculations \cite{KD04}.  It was shown in Refs.\ \cite{KD04,DD09} that no narrow resonances are predicted in the $j=1^-$ and $j=3^-$ partial waves.  All states in these partial waves therefore correspond to approximations of the continuum.

Notice that I only include $\he$ states with natural parity $(-)^j$. Other partial waves ($0^-,1^+,2^-,3^+$) are not directly
coupled to the ground state. They can be coupled only to partial waves with $j>0$, and are therefore expected to play a negligible role.
This will be discussed in the next subsection.

\subsection{Elastic cross sections}

I consider different systems where scattering data exist around the Coulomb barrier
($^{27}$Al, $^{58}$Ni, $^{120}$Sn, $^{208}$Pb).  These examples cover a wide range of target masses, and are investigated here within the same model and the same conditions of calculations.  The optical potentials $v_{pT}$ and $v_{nT}$ [see Eq.\ (\ref{eq10})] 
are taken from the compilation of Koning and Delaroche \cite{KD03}.  For the chosen targets, local potentials exist, and are specifically fitted to nucleon-target data.  This is in contrast with global potentials, whose parameters are fitted on different systems, and
then interpolated to the system considered.

For all calculations, I use an $R$-matrix channel radius $a = 24$ fm, with $N = 120$ mesh points.  The maximum angular momentum in the projectile target motion depends on the system and on the relative energy (typically $J_{\rm max}\sim 100-150$).  Many tests have been performed to check that all cross sections are numerically stable for small variations of these parameters.

\begin{figure}[htb]
	\begin{center}
		\epsfig{file=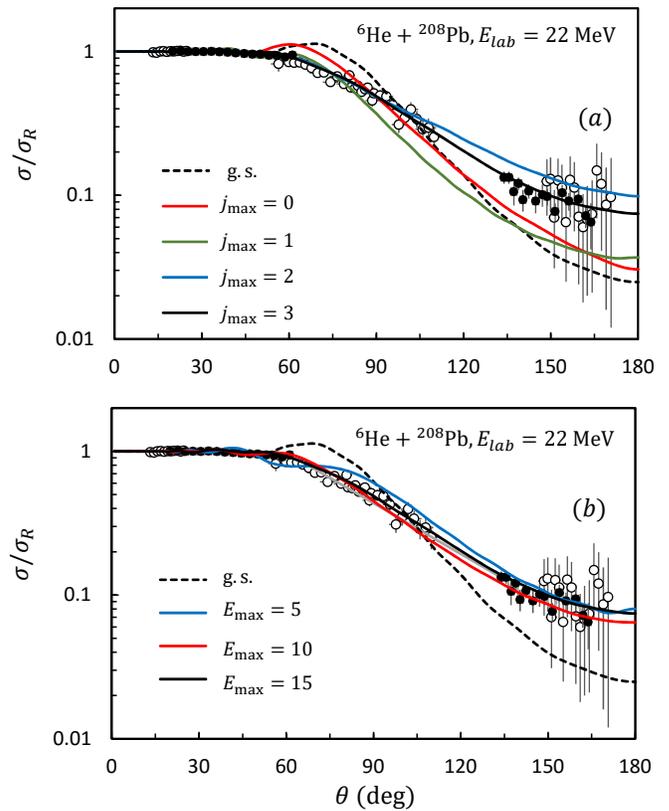,width=8.6cm}
		\caption{(Color online) Convergence of the $\hepb$ cross section at $E_{\rm lab}=22$ MeV, as a
			function of $j_{\rm max}$ (with $E_{\rm max}=15$ MeV) (a) and of $E_{\rm max}$ 
			(with $j_{\rm max}=3$) (b).
			The experimental data are taken from Ref.\ \cite{SEA08} 
			(full circles), and Ref.\ \cite{ASG11} (open circles).}
		\label{fig_conv}
	\end{center}
	
\end{figure}

I first analyze the convergence of the cross sections with the CDCC parameters $\jmax$ and $\emax$.  
This is presented in Fig.\ \ref{fig_conv} with the $\hepb$ system, at $\elab = 22$ MeV, where the convergence 
is the most critical.  Figure \ref{fig_conv}(a) illustrates the convergence with $\jmax$.  Clearly the 
single-channel approximation is unable to reproduce the data.  This was also observed in a 
non-microscopic approach \cite{RAG08}.  With a truncation energy $\emax = 15$ MeV, the $j = 0^+$ and 
$j = 1^-$ breakup contributions slightly improve the agreement between theory and experiment, but 
a fair agreement is obtained from $\jmax = 2$.  The convergence is excellent up to $\theta \approx 90^{\circ}$, but remains 
fair even at large backward angles.  The slowness of the convergence at energies close to 
the Coulomb barrier is well known \cite{RAG08}.

The lower panel of Fig.\ \ref{fig_conv} displays the convergence with $\emax$.  
Angular momenta up to $j=3$ are included.  Again the convergence is slow, and using the low 
truncation energy $\emax = 5$ MeV overestimates the cross section near $\theta \approx 90^{\circ}$. 
 As mentioned previously, 
this example is the most critical for convergence.  It is characterized by a heavy target, and by a 
low incident energy (the c.m. energy is $\ecm =21.4$ MeV, which is close to the Coulomb barrier $V_B=18.4$ MeV).  
For the other systems considered here, the convergence with $\jmax$ and $\emax$ is faster, and is not 
illustrated.

\begin{figure}[htb]
	\begin{center}
		\epsfig{file=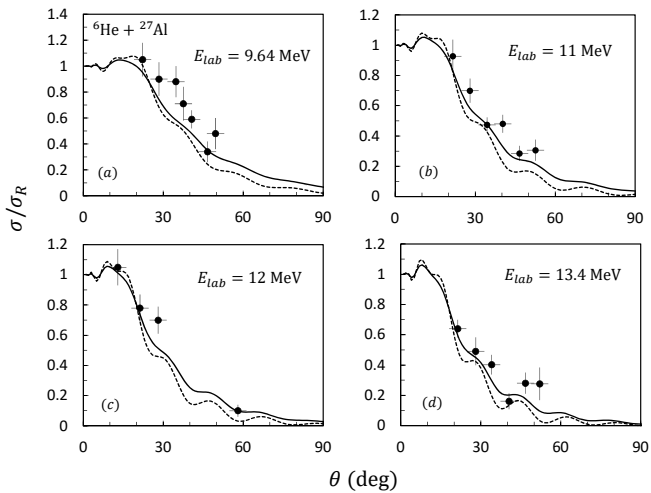,width=8.6cm}
		\caption{Elastic $\heal$ cross sections at different energies. The solid lines represent the full CDCC
			calculations, and the dotted lines represent the single-channel approximation. The
			data are taken from Ref.\ \cite{BLM07}.}
		\label{fig_secal}
	\end{center}
	
\end{figure}

\begin{figure}[htb]
	\begin{center}
		\epsfig{file=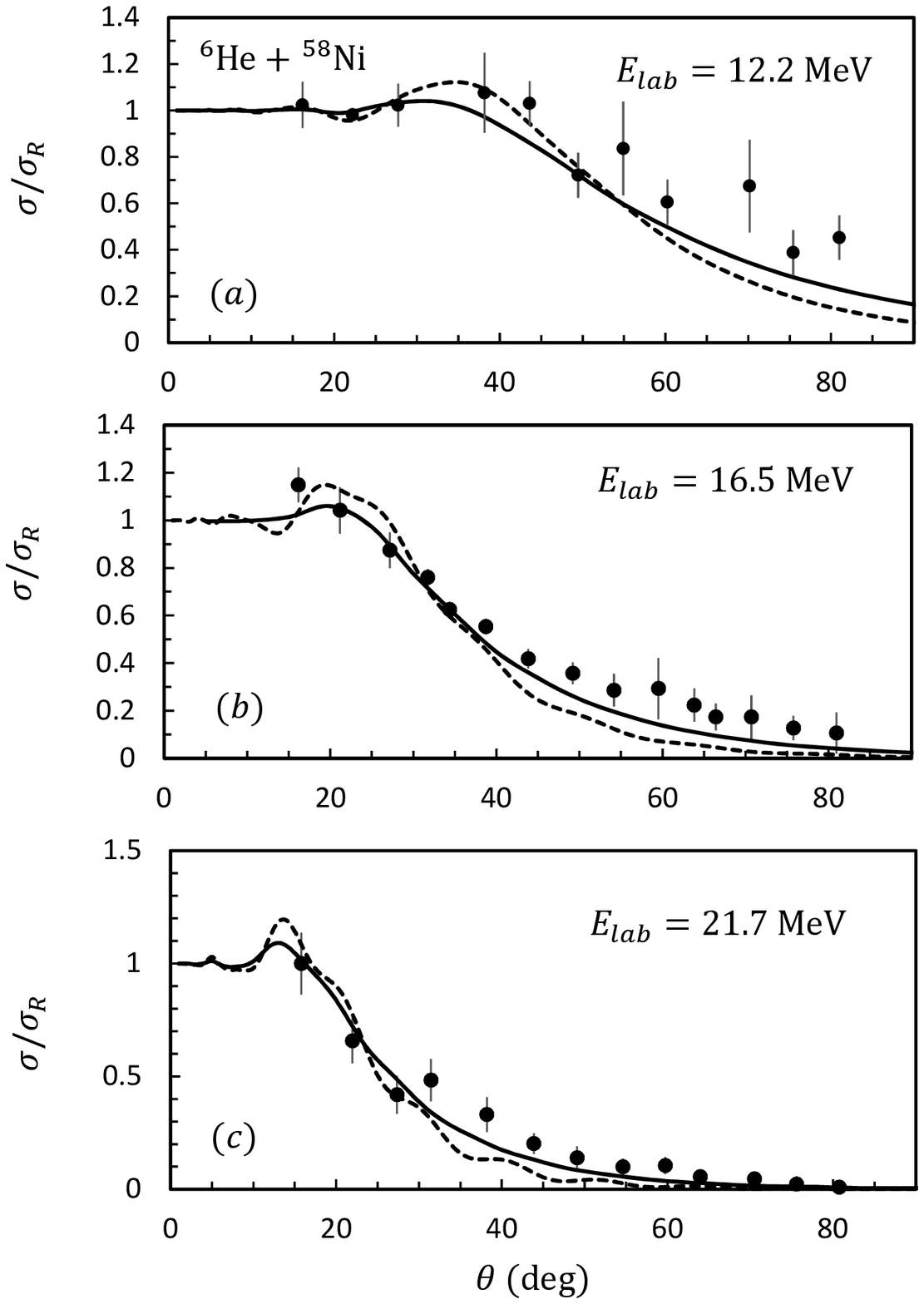,width=6.6cm}
		\caption{Elastic $\heni$ cross sections for the full CDCC calculation (solid line) and
			for the single-channel approximation. The experimental data are taken from Ref.\ \cite{MPR14}.}
		\label{fig_secni}
	\end{center}
\end{figure}

\begin{figure}[htb]
	\begin{center}
		\epsfig{file=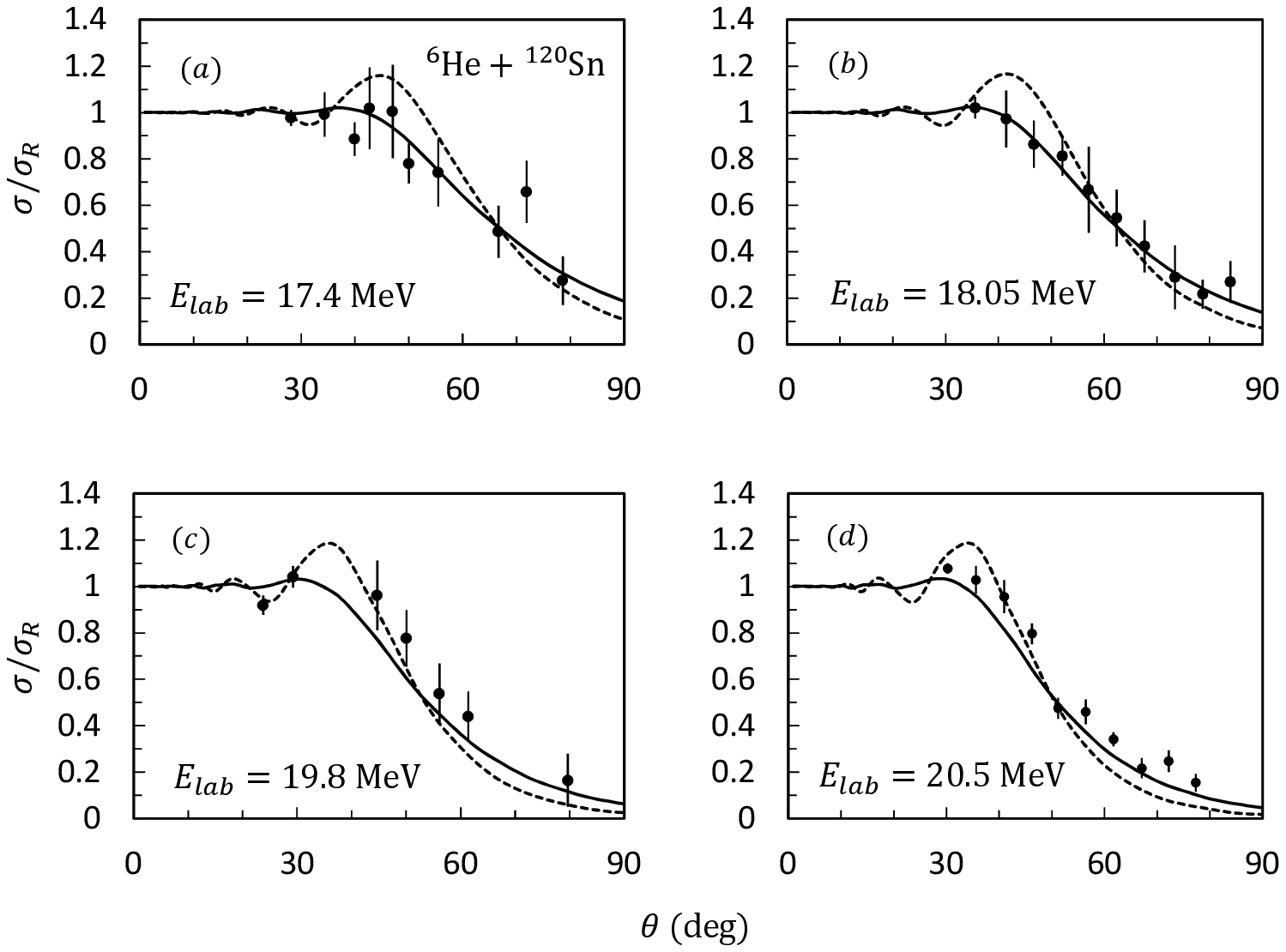,width=8.6cm}
		\caption{Elastic $\hesn$ cross sections for the full CDCC calculation (solid line) and
			for the single-channel approximation. The experimental data are taken from Ref.\ \cite{DLP10}.}
		\label{fig_secsn}
	\end{center}
\end{figure}

\begin{figure}[htb]
	\begin{center}
		\epsfig{file=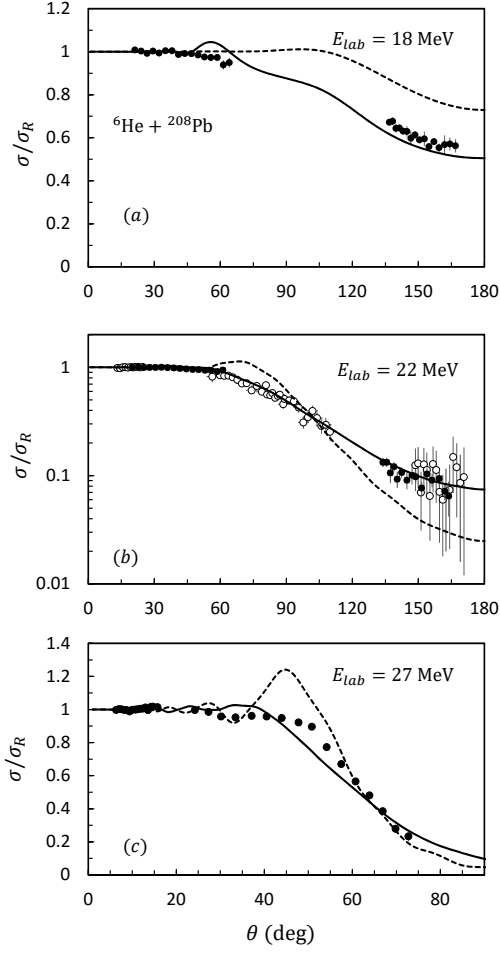,width=6.6cm}
		\caption{Elastic $\hepb$ cross sections for the full CDCC calculation (solid line) and
			for the single-channel approximation. The experimental data are taken from Ref.\ \cite{SEA08} (a),
			(b - full circles), Ref.\ \cite{ASG11} (b -  open circles), and Ref.\ \cite{KRS03} (c).}
		\label{fig_secpb}
	\end{center}
\end{figure}

In Figures \ref{fig_secal}, \ref{fig_secni}, \ref{fig_secsn}, and \ref{fig_secpb}, I present the 
CDCC cross sections for $^6$He scattering on $^{27}$Al, $^{58}$Ni, $^{120}$Sn and $^{208}$Pb, 
respectively.  These choices are guided by several reasons: $(i)$ covering a wide mass range, from light to heavy targets;
$(ii)$ experimental data are available; 
$(iii)$ local nucleon-target interactions, i.e. specifically fitted to nucleon scattering data, 
have been determined \cite{KD03}. 

Figure \ref{fig_secal} shows the $\heal$ system, at four energies.  These energies ($\ecm=7.8,
9.0,9.8,11.0$ MeV) 
are significantly higher than the Coulomb barrier ($V_B \approx 3.9$ MeV).  Although a 
slight improvement of the theoretical results is obtained within the multi-channel calculation, 
the single-channel approximation is not very different.  A similar conclusion has been drawn recently for 
the $^9$Be+$^{27}$Al system, in a non microscopic CDCC model \cite{CRA15}.  This weak sensitivity to 
breakup channels will be analyzed in more detail in Sec. IV.E.

In Figures \ref{fig_secni} and \ref{fig_secsn}, I present the $\heni$ and $\hesn$ cross sections, 
respectively.  Here the differences between the full calculation and the single-channel approximation 
are increasing.  The converged results are close to those of Ref.\ \cite{RAG08}, where a non-microscopic
$\alpha + n + n$ description of $\he$ was used.  In Fig.\ \ref{fig_secpb}, I consider the 
$\hepb$ system, which was used as an illustration of convergence issues in Fig.\ \ref{fig_conv}.  The 
strong influence of breakup channels is confirmed at the three energies.  Let us emphasize that 
all cross sections are obtained with the same conditions of calculations.  The only difference is 
that, of course, the choice of the neutron- and proton-target potentials is adapted to each system.

Finally, let us briefly discuss the role of non-natural-parity partial waves of $\he$ 
($j=0^-, 1^+, 2^-$, etc.).  These states cannot be coupled to the $j = 0^+$ ground state, 
but may play a role through couplings to the continuum.  As a full calculation, involving all pseudostates 
with $\jmax =3$ and $\emax = 15$ MeV is extremely demanding in terms of computer time and memory, I 
have performed two calculations with $\emax = 10$ MeV and $j=0^+, 1^-, 2^+$ or $j=0^{\pm}, 1^{\pm},2^{\pm}$. 
The difference between the two cross sections should provide a fair insight on the influence of 
non-natural-parity states.  The calculation has been done for $\hepb$ at $\elab = 22$ MeV. The differences
in the cross sections are, however, too small to be visible on a figure. The cross sections differ by less than $0.5\%$, and are therefore not shown.

\subsection{Role of the $\he$ halo in elastic scattering}
The role of a halo structure in nucleus-nucleus scattering takes its origin from the long range 
of the density in weakly bound nuclei.  Quantitatively, however, this influence of the halo is more 
difficult to assess.  In the present work, I investigate this effect by considering the short- 
and long-range parts of the nuclear densities.  For the $\he$ ground state, the proton and neutron GCM 
monopole densities shown in Fig.\ \ref{fig_dens} can 
be parametrized as
\beq
\rho_p(r)\approx \rho_{0p}\exp \bigl[-\bigl(\frac{r}{a_p}\bigr)^2\bigr],
\label{eq_rhop}
\eeq
with $\rho_{0p}=0.453 {\rm \ fm}^{-3}$ and $a_p=1.407$ fm, and by
\beq
\rho_n(r)\approx \rho_{0n}\biggl( \exp \bigl[ -\bigl(\frac{r}{a_n}\bigr)^2 \bigr] + \frac{0.02}{1+\exp(\frac{r-3.76}{0.8})}\biggr),
\label{eq_rhon}
\eeq
with $\rho_{0n}=0.426 {\rm \ fm}^{-3}$ and $a_n=1.690$ fm.
According to Eqs.\ (\ref{eq8a},\ref{eq8b}), these densities are normalized as
\beq
&& \int \rho_p (r) r^2 dr=2/\sqrt{4\pi}, \nonumber \\
&& \int \rho_n (r) r^2 dr=4/\sqrt{4\pi}.
\label{eq_norm}
\eeq
These approximations reproduce the exact calculations by less than $1\%$.  With the 
approximation (\ref{eq_rhon}), I can isolate the contribution from the core (first term) and the long-range part, 
associated with the halo component (second term).  

\begin{figure}[htb]
	\begin{center}
		\epsfig{file=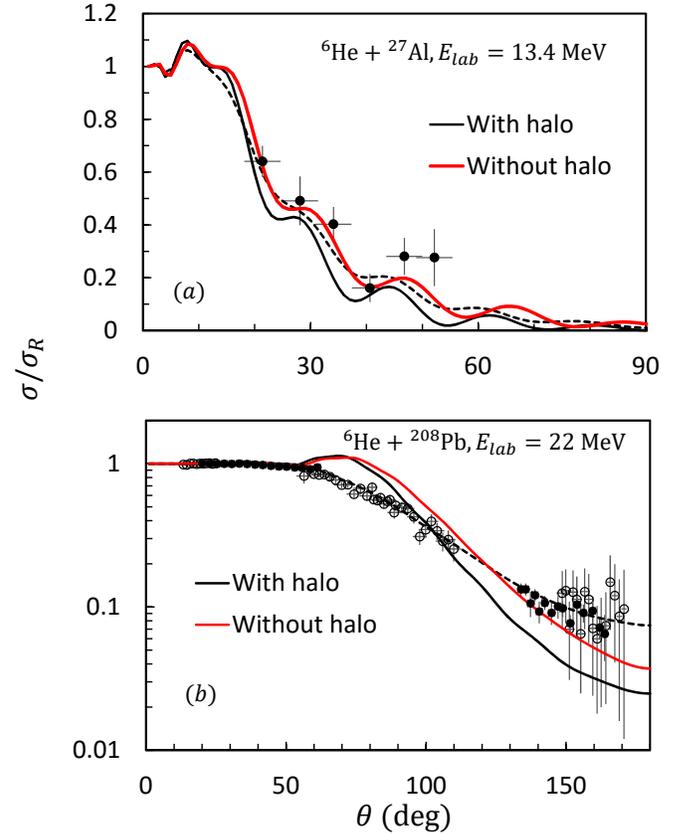,width=8.6cm}
		\caption{(Color online). Elastic $\heal$ (a) and $\hepb$ (b) cross sections with and without the halo component
			in the $\he$ neutron density. The dashed lines represent the full CDCC calculations. Experimental
			data are as in Fig.\ \ref{fig_secal} (a) and \ref{fig_secpb} (b).}
		\label{fig_halo}
	\end{center}
\end{figure}

In Fig.\ \ref{fig_halo}, I present calculations for the 
$^{27}$Al and $^{208}$Pb targets, either by including to core component only, or by including the 
full density.  In both cases, a single-channel calculation is performed, in order to isolate 
halo effects from breakup effects.  In both systems, the difference is small, in particular for $\heal$. 
Figure \ref{fig_halo} shows, with the $\hepb$ system, that breakup effects, obtained with the 
full continuum, are more important than halo effects.  This weak halo effect can be explained
by the small differences in the folding potentials.  A simple property of folding potentials 
is related to the volume integrals as 
\beq
\int V(\pmb{r})\, d\pmb{r}=A\int v(\pmb{r})\, d\pmb{r},
\label{eq_fold}
\eeq
where $v(\pmb{r})$ is the nucleon-target interaction (this identity holds for protons and neutrons 
separately). In other words, changing the density does not affect the volume integral.  The halo 
component of the neutron density (\ref{eq_rhon}) therefore modifies the range of the $^6$He-target 
potentials, but this effect is weak, as observed in the cross sections.

\subsection{Discussion of the $^6$He-target interaction}

The present model, including breakup channels, offers the possibility to analyze equivalent 
$^6$He-target potentials. Similar studies have been done for other projectiles, in
non-microscopic CDCC approaches (see, for example, Refs. \cite{MK04,MHO04,LN07,KKR13}).  

For a given partial wave $J\pi$, the equation associated with the elastic channel ise written as
\beq
&&\biggl( T_L(R)+V^{J\pi}_{11}(R)-E \biggr) g^{J\pi}_{1}(R)=\nonumber \\
&&-\sum_{c>1} V^{J\pi}_{1c}(R)g^{J\pi}_{c}(R).
\label{eq_pol1}
\eeq
A polarization potential $V_{pol}$  can be defined from
\beq
\biggl( T_L(R)+V^{J\pi}_{11}(R)+V^{J\pi}_{pol}(R)-E \biggr) g^{J\pi}_{1}(R)=0,
\label{eq_pol2}
\eeq
where 
\beq
V^{J\pi}_{pol}(R)=-\frac{\sum_{c>1} V^{J\pi}_{1c}(R)g^{J\pi}_{c}(R)}{g^{J\pi}_{1}(R)}.
\label{eq_pol3}
\eeq
With this definition, Eqs.\ (\ref{eq_pol1}) and (\ref{eq_pol2}) are strictly equivalent.  
However, the polarization potential (\ref{eq_pol3}) presents two disadvantages: $(i)$ it depends 
on $J$ and $\pi$; $(ii)$ it presents singularities at the nodes of the wave function.  

Thompson {\sl et al.} \cite{TNL89} proposed  an approximate, $J$ independent, polarization potential as
\beq
V_{pol}(R)=-\frac{\sum_{J\pi} V^{J\pi}_{pol}(R)\omega^{J\pi}(R)}{\sum_{J\pi}\omega^{J\pi}(R)},
\label{eq_pol4}
\eeq
where $\omega^{J\pi}(R)$ is a weight function, chosen as
\beq
\omega^{J\pi}(R)=(2J+1)(1-\vert U^{J\pi}_{11} \vert^2 ) \vert  g^{J\pi}_{1}(R) \vert^2.
\label{eq_pol5}
\eeq
This choice permits to avoid singularities in the potential, and to givet more weight on the important
partial waves (where $\vert U^{J\pi}_{11} \vert \ll 1 $).  It has been abundantly used in the 
literature (see references in Ref.\ \cite{CGD15}).  The reliability of the approximation can be tested by comparing the cross 
sections obtained with (\ref{eq_pol2}) and with the original CDCC calculation.

\begin{figure}[htb]
	\begin{center}
		\epsfig{file=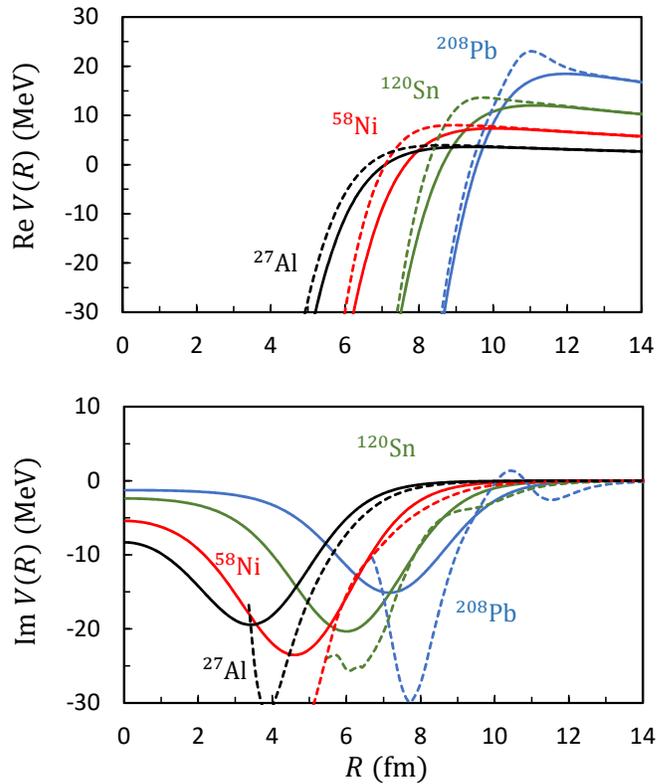,width=8.6cm}
		\caption{(Color online). Single-channel potentials $V_{11}(R)$ (solid lines) and total potentials $V_{11}(R)+V_{pol}(R)$ (dashed lines). The upper panel (a) represents the real part, and the 
			lower panel (b)	the imaginary part.}
		\label{fig_vpol}
	\end{center}
\end{figure}

In Fig.\ \ref{fig_vpol}, I present the total potential for the four systems considered here.  
Those potentials are determined at typical energies ($\ecm =11.0,11.1,16.6,21.4$ MeV for $\heal$, 
$\heni$, $\hesn$ and $\hepb$, respectively).  The general trend
is that the polarization potential increases for heavy systems.  This is consistent with the conclusions
drawn from the cross sections: breakup effects are weak for light targets, and increase for heavier 
targets.  The real part of the polarization potential is always repulsive, as usually 
observed \cite{CGD15}.  The imaginary part has a long tail, and is responsible for the long-range 
absorption.  For the $^{208}$Pb target, the imaginary part is negative beyond $R = 11$ fm, but presents 
a shape different from other targets.  This behaviour is expected from the strong breakup effects found 
with $^{208}$Pb.  

The accuracy  of the polarization potential has been tested by repeating the
calculation for many other numerical conditions (channel radius, number of basis functions, 
truncation energy and angular momentum).  For all reasonable choices of these parameters, 
the potentials are indistinguishable at the scale of the figure.

\subsection{Discussion of $\he$ breakup effects}
\begin{figure}[htb]
	\begin{center}
		\epsfig{file=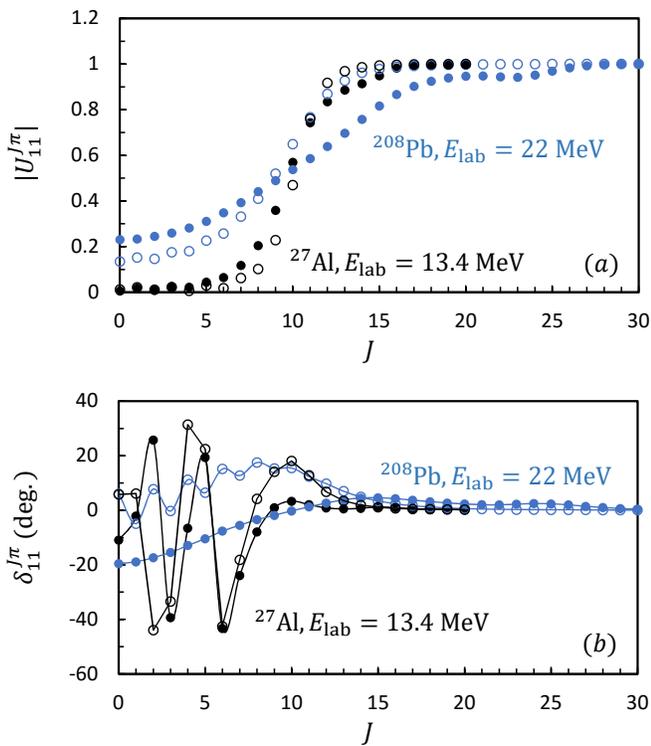,width=8.6cm}
		\caption{(Color online). Modulus (a) and phase (b) of the scattering matrix. Filled circles
			represent the full CDCC calculations, and the open circles represent the single-channel
			approximation. In (b) the lines are to guide the eye.}
		\label{fig_smat}
	\end{center}
\end{figure}

I showed in the previous subsections that breakup effects are weak for $^{27}$Al, and increase for 
heavier targets.  This effect can be traced in the scattering matrices $U^{J\pi}_{11}$.  For a given energy, 
the set of scattering matrices contains the same information as the elastic cross section. 
Figure \ref{fig_smat} displays the scattering matrices for the $^{27}$Al and $^{208}$Pb targets, and 
for the single-channel and full calculations.  They are parametrized as
\beq
U^{J\pi}_{11}=\vert U^{J\pi}_{11} \vert \exp(2i\delta^{J\pi}_{11}).
\eeq

For $^{27}$Al and $^{208}$Pb, the shape are clearly different.  Whereas low $J$ values 
are completely absorbed by the $^{27}$Al target ($ U^{J\pi}\simeq 0$ for $J \le 5$), they are still 
partly reflected in the $^{208}$Pb target.  As a general statement, this kind of figure presents 
three regions:
\begin{align}
\label{eq_bu1}
&\vert U^{J\pi}_{11} \vert \approx 0  {\rm \ for \ low \ } J {\rm \ values},\nonumber \\
&\vert U^{J\pi}_{11} \vert \approx 1  {\rm \ for \ high \ } J {\rm \ values},\nonumber \\
&0 \le \vert U^{J\pi}_{11} \vert \le  1  {\rm \ for\ intermediate \ } J {\rm \ values}. 
\end{align}

The precise shape of the nuclear potential does not affect neither region 1, nor region 2.  Figure  \ref{fig_smat}
shows that region 3, where the scattering matrices are sensitive to the interaction, is much wider 
for $^{208}$Pb than for $^{27}$Al.

In the limit of the sharp-absorption model, essentially developed by Frahn  \cite{Fr66} 
(see also Refs.\ \cite{Br85,CH13}), it is assumed that the transition occurs at a grazing angular momentum $J_g$.  In other words, 
I have, within this approximation
\begin{align}
\label{eq_bu2}
&\vert U^{J\pi}_{11} \vert= 0  {\rm \ for \ } J\le J_g,\nonumber \\
&\vert U^{J\pi}_{11} \vert= 1  {\rm \ for \ } J > J_g.
\end{align}
The model also assumes $(i)$ that the phase shifts $\delta^{J\pi}_{11}$ are zero, $(ii)$ that summations over 
the angular momentum can be replaced by integrals.  Even though this model has been essentially 
developed with the aim of investigating heavy-ion scattering at high energies, it remains valid 
provided the assumptions are satisfied.  Our goal here is not to use Frahn's model as a fit of 
the data, but to provide a simple estimate of the cross sections.  

Under these conditions, the scattering cross section is given by
\beq
\frac{d\sigma}{d\Omega}/\biggl( \frac{d\sigma}{d\Omega} \biggr)_R=
\frac{1}{2}\biggl[ (\frac{1}{2}-C(w))^2+(\frac{1}{2}-S(w))^2\biggr],
\label{eq_bu3}
\eeq
where $w$ is related to the scattering angle by
\begin{align}
\label{eq_bu4}
& w=\biggl[ \frac{J_g}{\pi \sin \theta_g}\biggr] ^{1/2}(\theta-\theta_g), \nonumber \\
& \theta_g=2 \arctan  \biggl( \frac{\eta}{J_g} \biggr),
\end{align}
where $\eta$ is the Sommerfeld parameter.

In (\ref{eq_bu3}), $C(w)$ and $S(w)$ are the Fresnel integrals.  This simple expression can be generalized 
to a smooth variation of the scattering matrix (\ref{eq_bu2}) around $J_g$ \cite{Fr66,Br85}.  Although more physical,
this extension leads to cross sections more complicated than (\ref{eq_bu3}).  As already mentioned, 
our aim is to provide a simple interpretation of the CDCC cross sections, and not perform optimal 
fits of the data.

\begin{figure}[htb]
	\begin{center}
		\epsfig{file=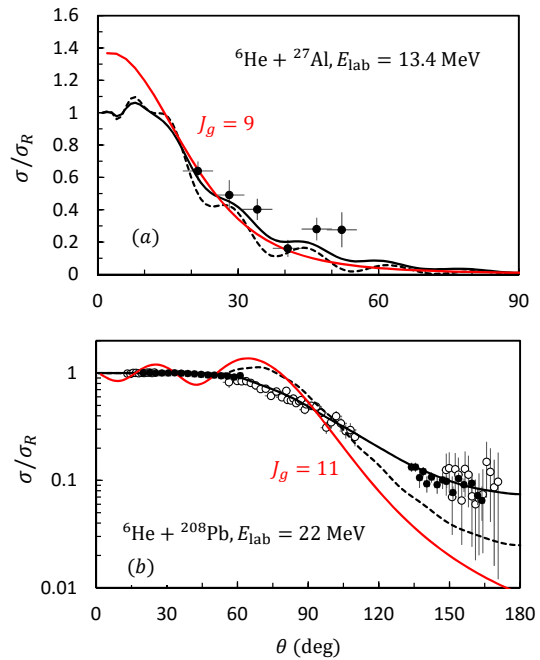,width=7cm}
		\caption{(Color online). Scattering cross sections computed with Eq.\ (\ref{eq_bu3}) (red lines)
			for the $\heal$ (a) and $\hepb$ (b) systems.
			The black solid lines represent the full CDCC calculations, and the dotted lines represent the single-channel approximation.
			}
		\label{fig_fresnel}
	\end{center}
\end{figure}

In Fig.\ \ref{fig_fresnel}, I compare approximation (\ref{eq_bu3}) with the CDCC calculations, for the $^{27}$Al and 
$^{208}$Pb targets.  The value of $J_g$ is estimated from Fig.\ \ref{fig_smat} ($J_g =9$ for $^{27}$Al, and
$J_g =11$ for $^{208}$Pb).  

Let us first discuss the  $^{27}$Al target, where differences between 
the single-channel and the full calculations are found fairly weak.  Here the Frahn approximation 
(\ref{eq_bu3}) provides a good description of the data and, except for $\theta \lesssim 20^{\circ}$, 
is in reasonable agreement with CDCC.  This is consistent with the scattering-matrix distribution, 
and explains why breakup effects play a minor role.  The data on $\heal$ scattering are weakly 
sensitive to the potential.  The only important parameter is the grazing angular momentum $J_g$ which 
is associated with the range of the imaginary potential \cite{Br85}.

The situation is different for the $\hepb$ system.  Here, approximation (\ref{eq_bu3}) is 
rather poor, as expected from the scattering matrices of Fig.\ \ref{fig_smat}.  Therefore the experimental 
cross section cannot be estimated from a simple model, and is more sensitive to the optical potential, or,
in other words, to the inclusion of breakup channels in the calculation.

\section{Conclusion}
\label{sec5}
In this work, I have applied the MCDCC method to the $\he$ nucleus, considered as an $\alpha+n+n$ three-cluster
system. The theory initiated in Ref.\ \cite{DH13} for two-cluster projectiles was extended to three-cluster nuclei. Although the main principles are identical, the numerical treatment of microscopic three-cluster
projectiles is much more involved. The $\he$ system is typical of Borromean nuclei, and is relatively simple
since the core is a $(0s)^4$ wave function. The conclusions drawn can be probably extended to other nuclei such as $^{11}$Li or $^{14}$Be, more demanding in terms of computer times since the core ($^{9}$Li or $^{12}$Be) involves $p$-shell orbitals.

The main advantage of the MCDCC is that it only relies on nucleon-target optical potentials, which are
in general well known. I have considered four different targets, $^{27}$Al, $^{58}$Ni, $^{120}$Sn, and
$^{208}$Pb, which should cover most typical masses. Around the Coulomb barrier, where experimental data are available, the elastic cross sections are fairly well reproduced by the model. In particular, the $\hepb$ data
at large backwards angles are sensitive to the conditions of the calculations. I have shown that breakup channels are crucial to explain the large experimental cross sections. As a general statement, I find that breakup effects are weak for light targets, and increase for heavier targets. Light systems have a low Coulomb barrier. In that case, either the energy is significantly larger (say 2 or 3 times the Coulomb barrier), and
the scattering matrices follow the sharp cut-off approximation, or the energy is around the Coulomb barrier, and most breakup channels are closed. As a consequence, data with light targets should be very accurate, and extend to
large backwards angles to be sensitive to the model.

This property is confirmed by an analysis of equivalent potentials. The polarization potential, induced by breakup effects, is small for light targets. The general trend is that polarization effects increase the Coulomb barrier, and provide a long-range absorption in the imaginary component of the nucleus-nucleus interaction.
The importance of breakup channels has been analyzed within the simple sharp cut-off approximation, where the scattering matrix is supposed to be either 0 below a grazing angular momentum, or 1 above this limit.
Even if this model is very basic, it provides a reasonable first guess of the physical
cross sections, and explains the weak breakup effects obtained for $^{27}$Al.

The present model could be generalized in various directions. Considering other projectiles, such as $^{11}$Li
or $^{14}$Be is a challenge for microscopic theories. The main limitation is the calculation of the GCM matrix 
elements (\ref{eq6}) which involve 7-dimension integrals. If the computer times remain within reasonable limits 
for $\he$, the necessity of $p$-orbitals represents a huge increase in the computational issues.
Other aspects of $\he$ scattering, such as breakup or fusion cross sections, are certainly worth being
investigated, and represent future applications of the MCDCC.

\section*{Acknowledgments}
This text presents research results of the IAP programme P7/12 initiated by the Belgian-state 
Federal Services for Scientific, Technical and Cultural Affairs. 


%

\end{document}